# High Precision Calibration Pairs for Northern Lucky Imaging

Matthew B. James[1], Graeme L. White[2], Roderick R. Letchford[2], Stephen G. Bosi[1].

1) University of New England, NSW, Australia.
   m.b.james27@gmail.com; sbosi@une.edu.au
2) University of Southern Queensland, QLD, Australia.
   graemewhiteau@gmail.com; Rod.Letchford@usq.edu.au

**Abstract.**

Presented here is list of 50 pairs quasi-evenly spaced over the northern sky, and that have Separations and Position Angles accurate at the milli-arcsec, and milli-degree level. These pairs are suggested as calibration pairs for lucky imaging observations. This paper is a follow-up to our previous paper regarding southern sky calibration pairs.

## 1. Introduction.

This paper is the second and final that lists low-relative proper motion pairs for lucky imaging calibration. The first paper, James *et al*., 2020c (under review) that looks at lucky imaging calibration pairs in the southern sky and outlines observational considerations on the limitations on accuracy, the atmosphere, the optics and camera. We recommend this paper and our previous James *et al*., 2020c be used in conjunction.

## 2. Calibration Pairs.

Based on the success of reference pair calibration used in James *et al* (2019a,b; 2020a) which presented (and investigated) PA and $\rho$ measurements and their associated uncertainties, and James *et al* 2020b (under review) which investigated the quality of three common calibration techniques; reference pair calibration, video drift method, and the use of an aperture diffraction grating; we present a list of 50 pairs sourced from the Washington Double Star Catalog (WDS) identified having characteristics for use in high precision calibration for PA and $\rho$.

The distribution of the pairs on the northern sky is shown in Figure 1. In the declination range from the equator to $+30^\circ$ there are 25 pairs, between $+30^\circ$ and $+60^\circ$ there are 18, and north of $+60^\circ$ there are 7.

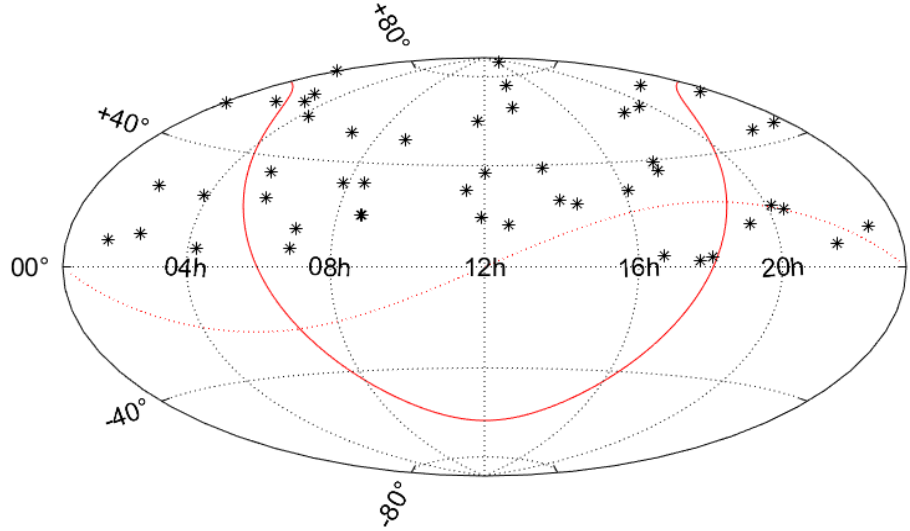

*Figure 1: Distribution of calibration pairs on the northern sky.*

Table 1 gives (i) the position of the primary star in J2000 coordinates, and (ii) the WDS code for the pair. The $\rho$ (iii) and PA (iv) are to allow identification 'on the night' as is the magnitudes of the primary (v) and the secondary (vi).

*Table 1: Positions and measures for the 50 northern pairs sourced from the WDS.*

| # | RA (J2000) | DEC (J2000) | DISC | Epoch | PA (deg) | ρ (arcsec) | Mag 1 (V) | Mag 2 (V) |
|---|---|---|---|---|---|---|---|---|
| 1 | 00 01 04.51 | +69 34 32.0 | STTA253AB | 2015 | 354 | 101.2 | 7.57 | 8.03 |
| 2 | 00 09 44.24 | +52 01 41.9 | BU  484AC | 2015 | 51 | 83.2 | 7.62 | 8.54 |
| 3 | 01 28 22.92 | +07 57 40.9 | S  398AB | 2018 | 100 | 68.9 | 6.34 | 8.02 |
| 4 | 01 52 40.77 | +57 17 17.4 | WAL  14AC | 2016 | 103 | 66.5 | 8.15 | 9.7 |
| 5 | 02 09 25.29 | +25 56 23.9 | H 6  69AC | 2014 | 279 | 106 | 5.02 | 7.97 |
| 6 | 02 26 45.64 | +10 33 55.0 | STTA 27AB | 2017 | 31 | 73.8 | 6.72 | 8.31 |
| 7 | 02 38 58.34 | +62 35 29.4 | STTA 28 | 2018 | 148 | 67.9 | 6.65 | 7.56 |
| 8 | 03 00 53.78 | +59 39 57.4 | STTA 31 | 2019 | 231 | 73.9 | 7.33 | 8.03 |
| 9 | 03 49 21.74 | +24 22 51.8 | STTA 40AB | 2018 | 308 | 86.7 | 6.58 | 7.53 |
| 10 | 04 15 28.86 | +06 11 13.6 | STTA 45AB | 2017 | 316 | 63.7 | 6.38 | 7.01 |
| 11 | 04 21 03.12 | +55 31 52.5 | STTA 46AB | 2017 | 160 | 99.9 | 7.68 | 7.95 |
| 12 | 05 25 30.00 | +34 11 10.9 | ARY  36 | 2015 | 246 | 110.4 | 6.75 | 8.48 |
| 13 | 05 47 56.13 | +24 41 12.6 | STTA 66 | 2015 | 166 | 93.7 | 6.95 | 7.71 |
| 14 | 06 54 06.05 | +06 41 15.8 | STTA 79 | 2015 | 89 | 115.9 | 7.2 | 7.52 |
| 15 | 06 58 06.59 | +14 13 43.4 | STTA 80AC | 2015 | 112 | 81.2 | 7.25 | 8.37 |
| 16 | 06 58 13.45 | +14 14 57.2 | STTA 80BC | 2015 | 193 | 106.8 | 7.37 | 8.37 |
| 17 | 06 53 30.93 | +51 38 42.4 | ES 2620AB | 2015 | 230 | 58.6 | 8.26 | 9.52 |
| 18 | 07 51 00.20 | +31 36 48.7 | STTA 89 | 2017 | 84 | 76.6 | 6.83 | 7.69 |
| 19 | 08 26 58.40 | +32 13 31.8 | HJL  93AB | 2015 | 193 | 107.3 | 8.43 | 9.25 |
| 20 | 08 39 50.72 | +19 32 26.9 | BU  584DC | 2015 | 89 | 99.5 | 6.67 | 7.47 |
| 21 | 08 40 22.11 | +19 40 11.9 | STF1254AC | 2018 | 343 | 62.5 | 6.52 | 7.61 |
| 22 | 09 05 39.04 | +50 17 41.7 | ES 2631 | 2017 | 259 | 78.9 | 7.84 | 8.47 |
| 23 | 11 41 25.80 | +58 55 26.0 | ES 2638 | 2015 | 352 | 30.5 | 8.93 | 9.5 |
| 24 | 11 30 04.32 | +29 57 52.7 | STTA111 | 2018 | 34 | 66.8 | 6.95 | 9.49 |
| 25 | 11 54 32.07 | +19 24 40.4 | STTA112AB | 2017 | 36 | 73.4 | 8.28 | 8.49 |
| 26 | 12 00 09.90 | +36 43 47.3 | STTA114 | 2018 | 83 | 86.3 | 7.48 | 8.39 |

| 27 | 12 39 09.20 | +16 28 47.2 | HJL1072 | 2015 | 23 | 118.3 | 7.63 | 9.26 |
|---|---|---|---|---|---|---|---|---|
| 28 | 13 27 04.70 | +64 44 07.6 | STTA123AB | 2017 | 145 | 69.2 | 6.65 | 7.03 |
| 29 | 13 46 59.77 | +38 32 33.7 | S 654AB | 2018 | 239 | 71.3 | 5.62 | 8.91 |
| 30 | 13 49 32.70 | +75 34 01.2 | SKF1229AB,C | 2015 | 199 | 99.9 | 7.65 | 9.19 |
| 31 | 14 04 45.95 | +25 49 03.9 | BGH 50 | 2015 | 32 | 97 | 7 | 8.9 |
| 32 | 14 34 20.57 | +24 23 30.5 | STTA129 | 2015 | 68 | 77.7 | 8.43 | 8.53 |
| 33 | 16 07 38.27 | +28 59 43.6 | GUI 16AC | 2015 | 350 | 118.1 | 7.93 | 9.48 |
| 34 | 16 40 38.69 | +04 13 11.3 | STFA 31AB | 2017 | 229 | 69.3 | 5.76 | 6.92 |
| 35 | 16 57 18.06 | +86 50 40.1 | STT 340AB | 2015 | 221 | 31.4 | 8.4 | 9.25 |
| 36 | 17 20 07.52 | +35 41 03.4 | SKF2822 | 2019 | 276 | 89.3 | 8.3 | 9.2 |
| 37 | 17 24 38.83 | +39 12 34.2 | S 689AB | 2017 | 197 | 89.2 | 7.48 | 8.44 |
| 38 | 17 39 08.48 | +02 01 41.2 | SHJ 251AB | 2015 | 328 | 111.3 | 6.37 | 7.78 |
| 39 | 18 01 32.36 | +03 31 27.4 | BU 1202AB,E | 2018 | 138 | 90.4 | 7.95 | 8.43 |
| 40 | 18 23 54.53 | +58 48 06.5 | STF2323BC | 2017 | 20 | 86.5 | 7.95 | 8.07 |
| 41 | 19 15 20.09 | +15 05 01.2 | STTA178 | 2018 | 267 | 89.6 | 5.69 | 7.64 |
| 42 | 19 33 10.07 | +60 09 31.3 | STFA 44 | 2017 | 287 | 74.9 | 6.47 | 8.19 |
| 43 | 20 09 55.48 | +21 00 07.3 | S 737 | 2015 | 128 | 101 | 7.93 | 9.26 |
| 44 | 20 30 14.44 | +19 25 15.7 | S 752AC | 2015 | 288 | 106.3 | 6.8 | 7.3 |
| 45 | 21 44 07.92 | +07 09 29.7 | STTA222 | 2018 | 258 | 87.4 | 7.49 | 8.47 |
| 46 | 21 57 11.10 | +66 09 22.2 | ARY 43 | 2010 | 128 | 100.4 | 6.38 | 6.75 |
| 47 | 22 09 15.19 | +44 50 47.3 | HJ 1735AD | 2015 | 286 | 110.2 | 6.73 | 6.77 |
| 48 | 22 58 33.60 | +12 03 04.4 | STTA241 | 2017 | 160 | 84.9 | 8.28 | 8.37 |
| 49 | 23 23 05.05 | +45 00 35.9 | SCA9004 | 2019 | 107 | 79.5 | 8.35 | 9.39 |
| 50 | 23 30 01.92 | +58 32 56.1 | SHJ 355AC | 2018 | 269 | 75 | 4.87 | 7.23 |

## 3.     Computation of Calibration $\rho$ and PA.

The GAIA DR2 catalogue (Brown, *et al.,* (2018)) was resourced for this work as the precision needed for PA and $\rho$ calibration needs to be much greater than measurements typically obtained by instruments used by astrometry community. GAIA DR2 reports position, proper motions and uncertainties with micro-arcsecond (µas) precision. The mean uncertainty of the 50 pairs presented here is ± 26 milli-arcseconds (mas) for $\rho$ and ± 3 milli-degree in PA at year 2100 (2000 to 2100).

Computation of PA and $\rho$ at an epoch of observation can be precisely estimated with the use of Equation 1 and the coefficients in Table 2 for each pair in this work. The linear equations of Equation 1 are justified as the selected pairs have both wide separations (>30") and have relatively (with respect to each component of the pair) low proper motions (~6 mas/yr). Practically, the change in PA and $\rho$ is linear over the period 2000 to 2100. Precession of PA over time has been accounted for.

The true change over time in PA and $\rho$ measures of wide pairs are better approximated with non-linear methods, but for a practically precise calibration this not needed. However, we also present non-linear equations (Appendix), and their coefficients, for lucky imaging calibration.

In Equation 1 the terms are:

   *$PA_t$*   The position angle, PA, of the pair on the night of observation *t*.
                    (*t* = date of observation – 2000.00).
   *A*       The annual increment to PA. From Table 2.
   *B*       The PA at Year 2000.00. From Table 2.
   *$\rho_t$*   The separation, r, of the pair on the night of observation *t*.
                    (*t* = date of observation – 2000.00).

| | C | The annual increment to ρ. From Table 2. |
|---|---|---|
| | D | The ρ at Year 2000.00. From Table 2. |

*Equation 1: Equations (i) PA and (ii) ρ at a user specified epoch, t, for pairs listed in Table 1.*

$$PA_t = At + B \qquad \& \qquad \rho_t = Ct + D$$

### 3.1. Example of Use

For the sake of illustration, lets us assume that the observations are being made on 21$^{st}$ April 2020, using the reference pair STTA253AB.

The Date (epoch) is $= 2020 + \dfrac{(31 + 29 + 31 + 21)}{365.25}$ (2020 is a leap year)

$t = 2020.307$

$t = 2020.307 - 2000$

$t = 20.307$

Obviously this calculation can be done in Excel and the day of the year (DOY) found on the web, using, for example, https://www.epochconverter.com/daynumbers.

Equation 1 now gives:

*Position Angle*

$$PA_{20.307} = -0.0058665(20.307) + 354.0613$$
$$= 353.941°$$

*Separation*

$$\rho_{20.307} = -0.0057667(20.307) + 101.15$$
$$= 101.033 \, arcsec$$

For the night of April 21, 2020, the PA of HJ 323 (RA = 00 01 04.51, Dec = +69 34 32.0; Mag 1 = 7.57, Mag 2 = 8.03) is 353.941°, and the separation is 101.033 arcsec. These values are accurate to ≈ 26 mas and ≈ 0.003°.

*Table 2: Table of coefficients used in Equation 1*

| | RA | DEC | Name | PA | | ρ | |
|---|---|---|---|---|---|---|---|
| | | | | A | B | C | D |
| 1 | 00 01 04.51 | +69 34 32.0 | STTA253AB | 5.8665E-03 | 3.5406E+02 | 5.7667E-03 | 1.0115E+02 |
| 2 | 00 09 44.24 | +52 01 41.9 | BU  484AC | -5.7363E-03 | 5.1488E+01 | 7.2958E-03 | 8.3205E+01 |
| 3 | 01 28 22.92 | +07 57 40.9 | S  398AB | 3.6402E-03 | 9.9539E+01 | 1.0613E-03 | 6.8994E+01 |
| 4 | 01 52 40.77 | +57 17 17.4 | WAL  14AC | 1.9967E-03 | 1.0271E+02 | 2.7930E-03 | 6.6528E+01 |
| 5 | 02 09 25.29 | +25 56 23.9 | H 6  69AC | -3.3584E-03 | 2.7901E+02 | 5.4191E-03 | 1.0581E+02 |
| 6 | 02 26 45.64 | +10 33 55.0 | STTA 27AB | 7.9110E-03 | 3.1890E+01 | -6.0335E-04 | 7.3637E+01 |
| 7 | 02 38 58.34 | +62 35 29.4 | STTA 28 | 6.4991E-03 | 1.4794E+02 | -1.7759E-03 | 6.7822E+01 |
| 8 | 03 00 53.78 | +59 39 57.4 | STTA 31 | 5.5091E-03 | 2.3030E+02 | 1.3108E-03 | 7.3638E+01 |
| 9 | 03 49 21.74 | +24 22 51.8 | STTA 40AB | 6.2800E-03 | 3.0894E+02 | -2.6733E-04 | 8.6916E+01 |

| 10 | 04 15 28.86 | +06 11 13.6 | STTA 45AB | 1.1158E-02 | 3.1570E+02 | -8.8407E-03 | 6.4342E+01 |
|---|---|---|---|---|---|---|---|
| 11 | 04 21 03.12 | +55 31 52.5 | STTA 46AB | -1.4753E-03 | 1.5984E+02 | 7.1286E-03 | 9.9707E+01 |
| 12 | 05 25 30.00 | +34 11 10.9 | ARY 36 | 6.6884E-03 | 2.4620E+02 | 2.9159E-04 | 1.1065E+02 |
| 13 | 05 47 56.13 | +24 41 12.6 | STTA 66 | 3.5499E-03 | 1.6613E+02 | -2.4279E-03 | 9.3822E+01 |
| 14 | 06 54 06.05 | +06 41 15.8 | STTA 79 | 3.8480E-03 | 8.9525E+01 | -9.1512E-04 | 1.1594E+02 |
| 15 | 06 58 06.59 | +14 13 43.4 | STTA 80AC | 2.0794E-03 | 1.1180E+02 | 1.7641E-03 | 8.1192E+01 |
| 16 | 06 58 13.45 | +14 14 57.2 | STTA 80BC | 6.6387E-03 | 1.9318E+02 | -4.3758E-03 | 1.0683E+02 |
| 17 | 06 53 30.93 | +51 38 42.4 | ES 2620AB | 1.2060E-02 | 2.3032E+02 | -5.1844E-03 | 5.8639E+01 |
| 18 | 07 51 00.20 | +31 36 48.7 | STTA 89 | 5.3657E-03 | 8.3496E+01 | -1.5560E-04 | 7.6827E+01 |
| 19 | 08 26 58.40 | +32 13 31.8 | HJL 93AB | -1.0341E-02 | 1.9347E+02 | 1.0675E-02 | 1.0722E+02 |
| 20 | 08 39 50.72 | +19 32 26.9 | BU 584DC | 4.0952E-03 | 8.8621E+01 | -4.1184E-04 | 9.9744E+01 |
| 21 | 08 40 22.11 | +19 40 11.9 | STF1254AC | 6.6133E-03 | 3.4264E+02 | 9.3609E-04 | 6.3356E+01 |
| 22 | 09 05 39.04 | +50 17 41.7 | ES 2631 | -4.3213E-03 | 2.5865E+02 | 7.8260E-03 | 7.9485E+01 |
| 23 | 11 41 25.80 | +58 55 26.0 | ES 2638 | 3.3669E-04 | 3.5243E+02 | -6.9800E-04 | 3.0445E+01 |
| 24 | 11 30 04.32 | +29 57 52.7 | STTA111 | -2.2352E-04 | 3.3305E+01 | 7.2053E-04 | 6.7207E+01 |
| 25 | 11 54 32.07 | +19 24 40.4 | STTA112AB | -5.4585E-04 | 3.5500E+01 | 8.0394E-04 | 7.3396E+01 |
| 26 | 12 00 09.90 | +36 43 47.3 | STTA114 | 1.0933E-02 | 8.2428E+01 | -6.4071E-04 | 8.6859E+01 |
| 27 | 12 39 09.20 | +16 28 47.2 | HJL1072 | 6.1250E-03 | 2.2689E+01 | 1.0165E-02 | 1.1811E+02 |
| 28 | 13 27 04.70 | +64 44 07.6 | STTA123AB | -2.5640E-03 | 1.4658E+02 | 5.1739E-04 | 6.8974E+01 |
| 29 | 13 46 59.77 | +38 32 33.7 | S 654AB | -9.9793E-03 | 2.3725E+02 | 1.4282E-03 | 7.1460E+01 |
| 30 | 13 49 32.70 | +75 34 01.2 | SKF1229AB,C | -4.4700E-03 | 1.9886E+02 | 2.2989E-03 | 9.9885E+01 |
| 31 | 14 04 45.95 | +25 49 03.9 | BGH 50 | -3.4221E-03 | 3.2056E+01 | 2.2256E-03 | 9.6944E+01 |
| 32 | 14 34 20.57 | +24 23 30.5 | STTA129 | -3.6574E-03 | 6.7798E+01 | -7.8664E-03 | 7.7764E+01 |
| 33 | 16 07 38.27 | +28 59 43.6 | GUI 16AC | -2.3858E-03 | 3.5033E+02 | -6.8296E-03 | 1.1827E+02 |
| 34 | 16 40 38.69 | +04 13 11.3 | STFA 31AB | -9.7048E-02 | 2.2942E+02 | -4.0957E-04 | 6.9660E+01 |
| 35 | 16 57 18.06 | +86 50 40.1 | STT 340AB | -7.5417E-03 | 2.2110E+02 | -1.1235E-03 | 3.1403E+01 |
| 36 | 17 20 07.52 | +35 41 03.4 | SKF2822 | -8.5829E-03 | 2.7636E+02 | -3.0230E-04 | 8.9338E+01 |
| 37 | 17 24 38.83 | +39 12 34.2 | S 689AB | -9.8709E-03 | 1.9652E+02 | -5.9539E-04 | 8.9687E+01 |
| 38 | 17 39 08.48 | +02 01 41.2 | SHJ 251AB | -5.7254E-03 | 3.2771E+02 | 4.9975E-04 | 1.1127E+02 |
| 39 | 18 01 32.36 | +03 31 27.4 | BU 1202AB,E | -1.2537E-02 | 1.3779E+02 | -8.3677E-04 | 9.0111E+01 |
| 40 | 18 23 54.53 | +58 48 06.5 | STF2323BC | 1.5843E-03 | 2.0781E+01 | -6.5176E-03 | 8.5757E+01 |
| 41 | 19 15 20.09 | +15 05 01.2 | STTA178 | -1.0602E-02 | 2.6705E+02 | 4.3164E-04 | 8.9826E+01 |
| 42 | 19 33 10.07 | +60 09 31.3 | STFA 44 | 7.8446E-04 | 2.8631E+02 | -9.4142E-03 | 7.5290E+01 |
| 43 | 20 09 55.48 | +21 00 07.3 | S 737 | -1.0793E-02 | 1.2811E+02 | 2.8474E-03 | 1.0096E+02 |
| 44 | 20 30 14.44 | +19 25 15.7 | S 752AC | -2.8443E-03 | 2.8796E+02 | -2.1840E-04 | 1.0628E+02 |
| 45 | 21 44 07.92 | +07 09 29.7 | STTA222 | -6.9733E-03 | 2.5742E+02 | -1.0214E-04 | 8.7630E+01 |
| 46 | 21 57 11.10 | +66 09 22.2 | ARY 43 | -4.6969E-03 | 1.2774E+02 | -2.6029E-03 | 1.0042E+02 |
| 47 | 22 09 15.19 | +44 50 47.3 | HJ 1735AD | -3.0684E-03 | 2.8569E+02 | 1.3367E-03 | 1.1001E+02 |
| 48 | 22 58 33.60 | +12 03 04.4 | STTA241 | -2.6307E-03 | 1.6072E+02 | -7.1155E-03 | 8.4786E+01 |
| 49 | 23 23 05.05 | +45 00 35.9 | SCA9004 | -1.3908E-03 | 1.0695E+02 | 2.0431E-04 | 7.9343E+01 |
| 50 | 23 30 01.92 | +58 32 56.1 | SHJ 355AC | -2.3152E-03 | 2.6875E+02 | -7.4036E-04 | 7.5628E+01 |

## 4.     Summary and Conclusion.

This work is the second and final presenting lucky imaging calibration pairs. The first paper presented 50 high precision calibration pairs for the southern hemisphere and discussed optical, atmospheric, camera and accuracy considerations for lucky imaging calibration, and discussed possible difficulties.

In this paper we presented 50 high precision calibration pairs quasi-evenly spaced across the northern sky for use with lucky imaging observations.

**Acknowledgements.**

- SIMBAD Astronomical Database, operated at CDS, Strasbourg, France, https://simbad.u-strasbg.fr/simbad/

- The Aladin sky atlas developed at CDS, Strasbourg Observatory, France, https://aladin.u-strasbg.fr/
- Brian D. Mason PhD & the Washington Double Star Catalog maintained by the UNSO. (WDS), https://ad.usno.navy.mil/wds
- The GAIA Catalogue (GAIA DR2, GAIA collaboration, 2018), from VizieR, http://vizier.u-strasbg.fr/viz-bin/VizieR-3?-source=I/345/gaia2

**References.**

Anton, R., (2014), "Double Star Measurements at the Southern Sky with 50 cm Reflectors and Fast CCD Cameras in 2012" *Journal of Double Star Observations*, **10**(3), 232-239.

Argyle, B., (2004), "Observing and Measuring Visual Double Stars", Springer press.

Brown, A.G.A., *et al.,* (2018), "Gaia Data Release 2 - Summary of the contents and survey properties", *Astronomy and Astrophysics*, **616**, A1.

Cox, A.N., (2002). *Allen's Astrophysical Quantities*., Springer, ISBN 978-1-4612-1186-0.

James, M.B., Emery, M., White. G.L., Letchford, R.R. & Bosi, S.G., (2019a), "Measures of Ten Sco Doubles and the Determination of Two Orbits." *Journal of Double Star Observations,* **15**(3), 489-503.

James, M.B., (2019b)., BSc Honours Thesis, University of New England, Australia.

James, M.B., Letchford, R.R., White, G.L., Emery, M., & Bosi, S.G., (2020a) "Measures of 62 Southern Pairs.", *Journal of Double Star Observations*, **16**(4), 325-348.

James, M.B., White, G.L., Bosi, S.B., & Letchford, R.R., (2020b (under review)), "Comparison of Image Scale Calibration Techniques: Known Pairs, Drift Scans and Aperture Grating", *Journal of Double Star Observations*, (Paper in Press).

James, M.B., White, G.L., Letchford, R.R., & Bosi, S.G., (2020c (under review)), "High Precision Calibration Pairs for Southern Lucky Imaging", *Journal of Double Star Observations*, (Paper in Press).

Prusti, T., *et al.,* (Gaia Collaboration), (2016), The Gaia mission, *Astronomy & Astrophysics*, **595**, A1.

White, G.L., Letchford, R.R. & Ernest, A.D., (2018). Uncertainties in Separation and Position Angle of Historic Measures – Alpha Centauri AB Case Study. *Journal of Double Star Observations*, **14**(3), 432.

**Appendix: Trigonometric and quadratic extrapolation of ρ and PA.**

Here we present inverse tangent and quadratic extrapolation for the above 50 calibration pairs. This is based on Equation A1, A2 and Table A1 and mimics the linear procedure above. Pairs denoted with an asterisk (*) require the use of Equation A1 (ii) to account for the trigonometric quadrants, and t is epoch of observation minus 2000. This procedure is also outlined in James *et al*., 2020c.

Equation A1: *Equations (i) & (ii) PA, and (iii) ρ at an observation epoch for pairs listed in Table 1.*

$$PA_t = tan^{-1}\left(\frac{At+B}{Ct+D}\right) \quad \& \quad PA_t^* = tan^{-1}\left(\frac{At+B}{Ct+D}\right) \pm 180 \quad \& \quad \rho_t = \sqrt{At^2 + Bt + C}$$

Equation A2: Precession *equation to account for precession in position angle.*

$$PA_p = PA_t - 0.0056 \sin(RA_{2000}) \sec(DEC_{2000}) (t)$$

*Table A1: PA and ρ coefficients used for Equation A1.*

| RA | DEC | Name | | PA | | | | ρ | | |
|---|---|---|---|---|---|---|---|---|---|---|
| | | | | A | B | C | D | A | B | C |
| 00 01 04.51 | +69 34 32.0 | STTA253AB | | 2.524E-06 | -2.909E-03 | 1.860E-06 | 2.795E-02 | 1.274E-04 | 1.157E+00 | 1.023E+04 |
| 00 09 44.24 | +52 01 41.9 | BU 484AC | | -4.208E-07 | 1.809E-02 | 3.748E-06 | 1.439E-02 | 1.844E-04 | 1.201E+00 | 6.923E+03 |
| 01 28 22.92 | +07 57 40.9 | S 398AB | * | 3.583E-07 | 1.890E-02 | 3.561E-07 | -3.176E-03 | 3.308E-06 | 1.462E-01 | 4.760E+03 |
| 01 52 40.77 | +57 17 17.4 | WAL 14AC | * | 8.503E-07 | 1.803E-02 | 2.458E-07 | -4.067E-03 | 1.015E-05 | 3.714E-01 | 4.426E+03 |
| 02 09 25.29 | +25 56 23.9 | H 6 69AC | | -2.011E-06 | -2.903E-02 | -3.202E-06 | 4.603E-03 | 1.853E-04 | 1.131E+00 | 1.120E+04 |
| 02 26 45.64 | +10 33 55.0 | STTA 27AB | | -3.556E-08 | 1.081E-02 | -1.753E-07 | 1.737E-02 | 4.145E-07 | -8.886E-02 | 5.422E+03 |
| 02 38 58.34 | +62 35 29.4 | STTA 28 | * | 1.075E-07 | 1.000E-02 | 6.500E-07 | -1.597E-02 | 5.625E-06 | -2.411E-01 | 4.600E+03 |
| 03 00 53.78 | +59 39 57.4 | STTA 31 | * | -3.625E-07 | -1.574E-02 | -1.333E-07 | -1.307E-02 | 1.933E-06 | 1.930E-01 | 5.422E+03 |
| 03 49 21.74 | +24 22 51.8 | STTA 40AB | | 3.892E-07 | -1.878E-02 | 3.625E-07 | 1.518E-02 | 3.666E-06 | -4.683E-02 | 7.554E+03 |
| 04 15 28.86 | +06 11 13.6 | STTA 45AB | | 2.205E-06 | -1.248E-02 | -1.282E-06 | 1.279E-02 | 8.428E-05 | -1.138E+00 | 4.140E+03 |
| 04 21 03.12 | +55 31 52.5 | STTA 46AB | * | 4.389E-06 | 9.547E-03 | -4.678E-07 | -2.600E-02 | 2.525E-04 | 1.401E+00 | 9.942E+03 |
| 05 25 30.00 | +34 11 10.9 | ARY 36 | * | -1.967E-07 | -2.812E-02 | 2.456E-07 | -1.240E-02 | 1.283E-06 | 6.441E-02 | 1.224E+04 |
| 05 47 56.13 | +24 41 12.6 | STTA 66 | * | 6.761E-07 | 6.246E-03 | 8.631E-07 | -2.530E-02 | 1.558E-05 | -4.566E-01 | 8.803E+03 |
| 06 54 06.05 | +06 41 15.8 | STTA 79 | | -2.636E-07 | 3.220E-02 | 9.608E-07 | 2.668E-04 | 1.287E-05 | -2.134E-01 | 1.344E+04 |
| 06 58 06.59 | +14 13 43.4 | STTA 80AC | * | 9.614E-07 | 2.094E-02 | 1.095E-06 | -8.376E-03 | 2.753E-05 | 2.840E-01 | 6.592E+03 |
| 06 58 13.45 | +14 14 57.2 | STTA 80BC | * | 1.329E-06 | -6.766E-03 | 9.392E-07 | -2.889E-02 | 3.433E-05 | -9.365E-01 | 1.141E+04 |
| 06 53 30.93 | +51 38 42.4 | ES 2620AB | * | -1.278E-08 | -1.254E-02 | 2.286E-06 | -1.040E-02 | 6.772E-05 | -6.121E-01 | 3.438E+03 |
| 07 51 00.20 | +31 36 48.7 | STTA 89 | | -3.917E-08 | 2.120E-02 | -3.806E-08 | 2.417E-03 | 3.865E-08 | -2.391E-02 | 5.902E+03 |
| 08 26 58.40 | +32 13 31.8 | HJL 93AB | * | 6.925E-06 | -6.936E-03 | -4.603E-06 | -2.896E-02 | 8.961E-04 | 2.211E+00 | 1.150E+04 |
| 08 39 50.72 | +19 32 26.9 | BU 584DC | | -1.194E-07 | 2.770E-02 | 2.050E-07 | 6.667E-04 | 7.295E-07 | -8.221E-02 | 9.949E+03 |
| 08 40 22.11 | +19 40 11.9 | STF1254AC | | 9.917E-08 | -5.251E-03 | 3.033E-07 | 1.680E-02 | 1.320E-06 | 1.186E-01 | 4.014E+03 |
| 09 05 39.04 | +50 17 41.7 | ES 2631 | * | -1.724E-06 | -2.165E-02 | -2.408E-06 | -4.346E-03 | 1.137E-04 | 1.239E+00 | 6.318E+03 |
| 11 41 25.80 | +58 55 26.0 | ES 2638 | | -4.750E-08 | -1.115E-03 | -2.019E-07 | 8.383E-03 | 5.578E-07 | -4.251E-02 | 9.269E+02 |
| 11 30 04.32 | +29 57 52.7 | STTA111 | | 1.056E-08 | 1.025E-02 | 2.325E-07 | 1.560E-02 | 7.020E-07 | 9.683E-02 | 4.517E+03 |
| 11 54 32.07 | +19 24 40.4 | STTA112AB | | -2.722E-08 | 1.184E-02 | 2.936E-07 | 1.660E-02 | 1.127E-06 | 1.180E-01 | 5.387E+03 |
| 12 00 09.90 | +36 43 47.3 | STTA114 | | 4.328E-07 | 2.392E-02 | -5.003E-06 | 3.179E-03 | 3.268E-04 | -1.439E-01 | 7.545E+03 |
| 12 39 09.20 | +16 28 47.2 | HJL1072 | | 6.908E-06 | 1.266E-02 | 1.064E-07 | 3.027E-02 | 6.187E-04 | 2.350E+00 | 1.395E+04 |
| 13 27 04.70 | +64 44 07.6 | STTA123AB | * | -1.000E-07 | 1.055E-02 | -2.381E-07 | -1.599E-02 | 8.640E-07 | 7.131E-02 | 4.757E+03 |
| 13 46 59.77 | +38 32 33.7 | S 654AB | * | -3.900E-07 | -1.670E-02 | -1.269E-07 | -1.074E-02 | 2.180E-06 | 2.041E-01 | 5.106E+03 |
| 13 49 32.70 | +75 34 01.2 | SKF1229AB,C | * | 3.750E-07 | -8.970E-03 | -8.022E-07 | -2.626E-02 | 1.016E-05 | 4.588E-01 | 9.977E+03 |
| 14 04 45.95 | +25 49 03.9 | BGH 50 | | 4.847E-07 | 1.429E-02 | 4.258E-07 | 2.282E-02 | 5.395E-06 | 4.315E-01 | 9.398E+03 |
| 14 34 20.57 | +24 23 30.5 | STTA129 | | -1.748E-06 | 2.000E-02 | -1.504E-06 | 8.162E-03 | 6.889E-05 | -1.224E+00 | 6.047E+03 |
| 16 07 38.27 | +28 59 43.6 | GUI 16AC | | 1.931E-06 | -5.516E-03 | -1.600E-06 | 3.239E-02 | 8.148E-05 | -1.619E+00 | 1.399E+04 |
| 16 40 38.69 | +04 13 11.3 | STFA 31AB | * | 3.056E-09 | -1.470E-02 | 1.714E-07 | -1.259E-02 | 3.808E-07 | -5.708E-02 | 4.853E+03 |
| 16 57 18.06 | +86 50 40.1 | STT 340AB | * | 2.953E-07 | -5.735E-03 | 1.567E-07 | -6.573E-03 | 1.448E-06 | -7.058E-02 | 9.862E+02 |
| 17 20 07.52 | +35 41 03.4 | SKF2822 | | 1.333E-08 | -2.466E-02 | -6.461E-07 | 2.748E-03 | 5.413E-06 | -5.455E-02 | 7.981E+03 |
| 17 24 38.83 | +39 12 34.2 | S 689AB | * | 1.849E-06 | -7.082E-03 | -3.683E-07 | -2.389E-02 | 4.606E-05 | -1.114E-01 | 8.044E+03 |
| 17 39 08.48 | +02 01 41.2 | SHJ 251AB | | -1.403E-07 | -1.651E-02 | 7.556E-08 | 2.613E-02 | 3.290E-07 | 1.112E-01 | 1.238E+04 |
| 18 01 32.36 | +03 31 27.4 | BU 1202AB,E | * | 4.386E-07 | 1.682E-02 | 7.133E-07 | -1.854E-02 | 9.088E-06 | -1.516E-01 | 8.120E+03 |
| 18 23 54.53 | +58 48 06.5 | STF2323BC | | 2.069E-06 | 8.452E-03 | -2.741E-06 | 2.227E-02 | 1.528E-04 | -1.129E+00 | 7.354E+03 |
| 19 15 20.09 | +15 05 01.2 | STTA178 | * | -1.125E-07 | -2.492E-02 | -1.458E-07 | -1.285E-03 | 4.397E-07 | 7.752E-02 | 8.069E+03 |

| | | | | | | | | | |
|---|---|---|---|---|---|---|---|---|---|
| 19 33 10.07 | +60 09 31.3 | STFA 44 | | 3.109E-06 | -2.007E-02 | 1.275E-06 | 5.872E-03 | 1.463E-04 | -1.423E+00 | 5.669E+03 |
| 20 09 55.48 | +21 00 07.3 | S 737 | * | -2.461E-06 | -2.207E-02 | -1.882E-06 | 1.731E-02 | 1.244E-04 | 5.633E-01 | 1.019E+04 |
| 20 30 14.44 | +19 25 15.7 | S  752AC | | 1.042E-07 | -2.808E-02 | 1.244E-07 | 9.105E-03 | 3.413E-07 | -4.645E-02 | 1.129E+04 |
| 21 44 07.92 | +07 09 29.7 | STTA222 | * | 2.194E-08 | -2.376E-02 | 3.194E-08 | -5.300E-03 | 1.947E-08 | -1.790E-02 | 7.679E+03 |
| 21 57 11.10 | +66 09 22.2 | ARY  43 | * | -2.611E-07 | 2.206E-02 | 8.447E-07 | -1.707E-02 | 1.013E-05 | -5.231E-01 | 1.008E+04 |
| 22 09 15.19 | +44 50 47.3 | HJ 1735AD | | -5.817E-07 | -2.942E-02 | -7.017E-07 | 8.266E-03 | 1.077E-05 | 2.932E-01 | 1.210E+04 |
| 22 58 33.60 | +12 03 04.4 | STTA241 | * | -1.269E-07 | 7.777E-03 | 2.050E-06 | -2.223E-02 | 5.469E-05 | -1.207E+00 | 7.189E+03 |
| 23 23 05.05 | +45 00 35.9 | SCA9004 | * | 5.417E-08 | 2.108E-02 | -1.694E-08 | -6.424E-03 | 4.175E-08 | 3.242E-02 | 6.295E+03 |
| 23 30 01.92 | +58 32 56.1 | SHJ 355AC | * | 2.258E-07 | -2.100E-02 | -8.433E-07 | -4.586E-04 | 9.878E-06 | -1.129E-01 | 5.720E+03 |